\documentclass[preprints,article,accept,moreauthors,pdftex]{mdpi} 
\usepackage{gensymb}
\usepackage[utf8]{inputenc}
\usepackage[T1]{fontenc}

\firstpage{1} 
\pubvolume{xx}
\issuenum{1}
\articlenumber{5}
\pubyear{2019}
\copyrightyear{2019}

 \history{Published: 16 February 2019 }

\Title{Engineering topological nodal line semimetals in Rashba spin-orbit coupled atomic chains }

\Author{Paola Gentile $^{1,2 *}$, Vittorio Benvenuto $^{2}$, Carmine Ortix  $^{2,3}$, Canio Noce $^{2}$ and Mario Cuoco $^{1,2 *}$}

\AuthorNames{Paola Gentile, Vittorio Benvenuto, Carmine Ortix, Canio Noce and Mario Cuoco}

\address{
$^{1}$ \quad CNR-SPIN, I-84084 - Fisciano (Salerno), Italy

$^{2}$ \quad Dipartimento di Fisica “E.R. Caianiello”, Universita` degli Studi di Salerno, Via G. Paolo II, 132 I-84084 Fisciano (Salerno), Italy  

$^{3}$ \quad Institute for Theoretical Physics, Center for Extreme Matter and Emergent Phenomena, Utrecht University, Princetonplein 5, 3584 CC Utrecht, The Netherlands }

\corres{Correspondence: paola.gentile@spin.cnr.it; mario.cuoco@spin.cnr.it; Tel.: +39 089 969149 (P.G.); +39 089 969328  (M.C.)}

\abstract{We study an atomic chain in the presence of modulated charge potential and modulated Rashba spin-orbit coupling (RSOC) of equal period. We show that for commensurate periodicities $\lambda=4 n$ with integer $n$, 
the three-dimensional synthetic space obtained by sliding the two phases of the charge potential and RSOC features a topological nodal line semimetal protected by an antiunitary particle-hole symmetry. The location and shape of the nodal lines strongly depends on the relative amplitude between the charge potential and RSOC.  }


\begin{document}



\section{Introduction}

The discovery of time-reversal invariant topological insulators \cite{art1}-\cite{art4} has triggered in the past few years a huge interest in these new quantum states of matter \cite{art5}-\cite{art12}. One of the distinguishing features of topological insulators is the existence of conducting boundary states within a bulk energy gap \cite{art9}.  The existence of these topologically protected states can be inferred from the knowledge of a topological quantity, such as the Chern number or the Z$_{2}$ index \cite{art13}-\cite{art15}, which is only determined by the topology of the electronic wavefunctions. 
On general grounds, one therefore expects that band structure engineering can be employed to possibly induce topological phase transition \cite{art1}-\cite{art2}.
Indeed, several approaches have been theoretically proposed to drive a topological phase transition using strong doping \cite{art17}, electric fields \cite{art18},\cite{art19} high pressure \cite{art20}, etc. 

Alternative proposals have demostrated the possibility to drive a trivial system into a topological insulator by introducing a superlattice structure. 
Generally speaking, the concept of superlattice introduced by Esaki and Tsu \cite{art21} has been widely employed as a powerful method to engineer the  electronic band structures of conventional semiconductors for various technological applications \cite{art22}-\cite{art24}. 
   
In the specific context of topological phase transitions, it has been demostrated that the application of a spatial periodic charge potential can drive a metallic system into a Chern insulator, characterized by gapped regions in the energy spectrum having non-zero Chern numbers \cite{PhysicaB}.
A spatial periodic charge potential has been also demostrated to turn a conventional insulator into a quantum spin Hall system. Considering the prototype case of HgTe$/$CdTe quantum wells, it has been demonstrated \cite{PRB2014} that at a critical strength of the charge potential, a conventional band insulator with strong spin-orbit coupling is driven into a quantum spin Hall system associated with band inversion and 
 consequently spin-momentum locked edge states. 

Insulating phases with a non-trivial topology can be also induced in one-dimensional atomic chains by a periodic canting of the Rashba spin-orbit field \cite{nostroPRL2015}. 
Infact, for certain corrugation periods the system possesses topologically non-trivial insulating phases at half-filling, with topologically protected zero-energy modes. 
  Relevantly, such a system, under the application of a rotating magnetic field, can realize the Thouless topological pumping protocol in an entirely novel fashion \cite{PRBserpentine}.  
  
Furthermore, when the sublattice structure of the charge potential and Rashba spin-orbit coupling is mirror point invariant, a class of one--dimensional time-reversal invariant insulators beyond the standard Altland-Zirnbauer classification can be realized \cite{OrtixLau2016}. 

The next step is to understand how the effects of a spatially periodic charge potential combine with that of a spatially modulated Rashba spin-orbit coupling in an atomic chain.   The interplay between these two modulated fields is expected to be highly non trivial, in particular from a symmetry perspective. Indeed, charge potential and RSOC satisfy different symmetries, and act on distinct electron degrees of freedom. The periodic charge potential behaves like a local source of chemical potential modulation, while the RSOC comes from inversion symmetry breaking and leads to electron spin and momentum locking. In consideration of the fundamental role played by symmetries in determining the existence and the nature of topological phases, the  simultaneous action of charge potential and RSOC, especially in the presence of additional space symmetries related to periodicity, can offer new possibilities to manipulate topological states and eventually induce novel topological phases. 
 
In this paper we address this problem, focusing in particular on the character of topological states emerging at half-filling in the case of periodicity $\lambda=4 n$, with integer $n$. 
We show that in the three-dimensional synthetic space obtained by sliding the phases of the charge potential and RSOC such periodicity leads at half-filling to a topological nodal line phase protected by particle-hole symmetry. The relative amplitude between the charge potential and RSOC strongly affects the shape of the nodal lines, also determining the regimes where the system is topologically trivial or topologically non trivial.    
After introducing the model Hamiltonian for the considered one--dimensional system, we show the effects of spatial periodicity in the charge potential and RSOC on the system energy spectrum, showing the appearance of systematic energy gap closing and reopening at half filling in the space defined by the slide phases of the two periodic fields.
We discuss the topological character of the emerging insulating phases in terms of the Hamiltonian symmetries and topological invariants. 
 
\section{The Model}
We consider a one dimensional system of spin one-half fermions in the presence of spatially modulated periodic charge potential $V(x)= V_0  Cos( 2 \pi x q_V+\phi_V)$ and a RSOC  described by a field which has periodic amplitude $\alpha_z(x)= \alpha_0 Cos(2 \pi q_{\alpha} x +\phi_{\alpha})$ and a fixed $\hat{z}$ direction.
We assume that the modulating periods of the two fields assume rational values and are equal, $\lambda = 1/q_V = 1/q_{\alpha}$. Moreover, $\phi_{\alpha}$ and $\phi_{V}$ are slide phases for the two considered periodic fields. 
 
  The Hamiltonian of such system in a lattice formalism is a generalization of the famous Aubry-Andr\'{e}- Harper (AAH) model \cite{AAH1}-\cite{AAH3}, with the inclusion of a spatially periodic RSOC: 
\begin{eqnarray}
\mathcal{H}=\mathcal{H}_0  +\mathcal{H}_V +\mathcal{H}_{RSOC}    \label{Hreal}
\end{eqnarray}
where
 
\begin{eqnarray}
 {\mathcal{H}}_{0} &=& -t \sum_{j=1}^{N-1} ( c^{\dag}_{j, \sigma}c_{j,n+1 \sigma}+ h.c. )-\tilde{t} \left(c^{\dag}_{N, \sigma} c_{1, \sigma}+c^{\dag}_{1, \sigma} c_{N, \sigma}  \right)\\
{\mathcal{H}}_{V} &=& \sum_{j=1}^N V(j) c^{\dag}_{j \sigma}c_{j \sigma} \\
{\mathcal{H}}_{RSOC} &=& i \left[\sum_{j=1}^{N-1} 
\alpha_z(l) c^{\dag}_{j, \sigma} s_z^{\sigma \sigma'} c_{j+1, \sigma'} 
- \sum_{j=2}^{N} \alpha_z(j-1) c^{\dag}_{j, \sigma} s_z^{\sigma \sigma'} c_{j-1, \sigma'}\right] \\ &&+ i \tilde{\alpha} \left( c^{\dag}_{N, \sigma}s_{z}^{\sigma \sigma'}c_{j, \sigma'}- c^{\dag}_{j, \sigma} s_{z}^{\sigma \sigma'} c_{N, \sigma} \right).
\end{eqnarray}

\noindent The operators $c_{j\sigma}^{\dag}(c_{j\sigma})$ create (annihilate) an electron with spin $\sigma \ (\sigma= \uparrow,\downarrow)$ at the lattice site $j$, $s_z$ is the Pauli matrix describing the spin operator along the $\hat{z}$ direction, $N$ is the number of total sites of the considered atomic chain and $t$ is the hopping amplitude between nearest neighbour sites. $\tilde{t}$ and $\tilde{\alpha}$ are the hopping amplitude and antisymmetric Rashba spin-orbit, respectively, between the last and the first site of the atomic chain, such that they both assume zero value for an open chain and are $\tilde{t}=t$ and $\tilde{\alpha}=\alpha_z(N)$ for closed boundary conditions.

The periodic variations of the charge potential and RSOC introduce a superlattice structure which allow to represent the position of each lattice site in terms of two indices: one ($R_I$) indicating the position of the supercell, and the other ($j$) representing the atomic position of the lattice site inside the specified supercell. As a consequence, the Hamiltonian of eq.(\ref{Hreal}) can be written down as:

\begin{eqnarray}
 {\mathcal{H}}_{0} &=& \sum_{\sigma} \left[
\sum_{R_I=1}^{N_s} \sum_{j=1}^{\lambda-1}
(-t) c^{\dag}_{j, R_I, \sigma}c_{j+1, R_I, \sigma} + \sum_{R_I=1}^{N_s-1} (-t)c^{\dag}_{\lambda, R_I, \sigma} c_{1,R_I +1, \sigma} \right]\\
&&+ \sum_{\sigma}( -\tilde{t}) \left(c^{\dag}_{\tilde{N}, N_s,\sigma} c_{1,1,\sigma}+c^{\dag}_{1, 1,\sigma} c_{\tilde{N}, N_s,\sigma} \right) + h. c.\\
 {\mathcal{H}}_{V} &=& \sum_{j=1}^{N_s} \sum_{R_I=1}^{\lambda} V(j) c^{\dag}_{j, R_I, \sigma} c_{j, R_I, \sigma} \\
 {\mathcal{H}}_{RSOC}& =& i \sum_{\sigma,\sigma'} \left \{
\left[ \sum_{R_I=1}^{N_s} \left( 
\sum_{j=1}^{\lambda-1}\alpha_z(j) c^{\dag}_{j, R_I, \sigma} s_z^{\sigma  \sigma'}c_{j+1, R_I, \sigma'} - \sum_{j=2}^{\lambda} \alpha_z(j-1) c^{\dag}_{j, R_I, \sigma}s_z^{\sigma\sigma'} c_{j-1, R_I, \sigma'} \right) 
\right] \right.\\
&+& \left. \left[\left( \alpha_z (\lambda) \sum_{R_I=1}^{N_s-1} c^{\dag}_{\lambda, R_I, \sigma} \sigma_{z}^{\sigma \sigma'} c_{1, R_I+1, \sigma'} + \tilde{\alpha} \   c^{\dag}_{\tilde{N}, N_s, \sigma}\sigma_{z}^{\sigma \sigma'} c_{1,1, \sigma'} \right)+ h.c.  \right]\right \} .
\end{eqnarray}

\noindent where $N_s=N/\lambda$ is the total number of supercells in the chain and $\tilde{N}$ is the last site of the last supercell in the chain. We will assume that the chain has a number of sites which is integer mutiple of the periodicity $\lambda$, such that $\tilde{N}=\lambda$. 

\noindent Then, in the case of closed boundary conditions ($\tilde{t}=t$ and $\tilde{\alpha}=\alpha_z(\lambda)$), we can Fourier transform the 
creation and annihilation operators $c^{\dag}_{j, R_I, \sigma},c_{j, R_I, \sigma}$
with respect the supercell index $R_I$:

\begin{eqnarray}
&& c^{\dag}_{j, R_I, \sigma}=\frac{1}{\sqrt{N_s}}\sum_{k}e^{i\ k\ R_{I}} c^{\dag}_{j,k,\sigma}\\
&& c_{j, R_I, \sigma}=\frac{1}{\sqrt{N_s}}\sum_{k}e^{-i \ k \ R_{I}} c_{j,k,\sigma}
\end{eqnarray}
\noindent After Fourier transforming, the Hamiltonian in momentum space reads as: \begin{eqnarray}
\mathcal{H}& =& \sum_{k}\psi_{k}^{\dag} H_k \psi_k  
\end{eqnarray}

\noindent with $\psi_{k}=(c_{1,k,\uparrow},..., c_{\lambda, k, \uparrow},c_{1,k,\downarrow},..., c_{\lambda,k,\downarrow})$ and  

\begin{eqnarray}
\nonumber  \mathcal{H}_k &=&
\sum_{k,\sigma}\sum_{j=1}^{\lambda-1} ( - t) \left[ \left( c^{\dag}_{j , k, \sigma} c_{j+1, k, \sigma }+ c^{\dag}_{j+1, k,\sigma} c_{j, k, \sigma} \right)  +   \left( c^{\dag}_{\lambda, k, \sigma} c_{1, k, \sigma}  e^{- i \ k \ \lambda \ a} + c^{\dag}_{1, k, \sigma} c_{\lambda, k,  \sigma}  e^{ i\  k\  \lambda \ a} \right) \right]  \\ 
\nonumber  &+& i  \sum_{k} \sum_{\sigma,\sigma'} \left[  \left( \sum_{j=1}^{\lambda-1}    \alpha_z(j) c^{\dag}_{j, k, \sigma} \sigma_z^{\sigma \sigma'} c_{j+1, k, \sigma'} \right) -  \left( \sum_{j=2}^{\lambda}  \alpha_z(j-1) c^{\dag}_{j, k, \sigma}  \sigma_z^{\sigma \sigma'} c_{j-1, k, \sigma'}  \right) \right. + 
\\
&+& \left. \alpha_z (\lambda) \sum_{k, \sigma, \sigma'} \left( c^{\dag}_{\lambda, k, \sigma} \sigma_{z}^{\sigma \sigma'} c_{1, k, \sigma'}   e^{- i \ k \ \lambda \ a} - c^{\dag}_{1, k, \sigma'}\sigma_z^{\sigma \sigma'}  c_{\lambda, k,  \sigma}  e^{ i\  k\  \lambda \ a} \right)\right]  \\ 
&+& \sum_{k,\sigma}\sum_{j=1}^{\lambda}  V(j) c^{\dag}_{j \ k \ \sigma} c_{j \ k\  \sigma}. 
\end{eqnarray} \label{Hk} 
 
\section{Metal-insulator transition induced by periodicity}

Due to the Abelian character of the RSOC, the Hamiltonian commutes with the spin operator $s_z$ and can be therefore brought in the block diagonal form for every periodicity $\lambda$:

\begin{eqnarray} 
\mathcal{H}_k= \left( 
\begin{array}{cc} 
   \mathcal{H}_{\uparrow \ k}             & 0 \\
   0 &      \mathcal{H}_{\downarrow \ k}  \\
\end{array}
\right),
\end{eqnarray}

\noindent where each sub--block $\mathcal{H}_{\sigma \ k}$ corresponds to one of the two electron spin projections $\sigma=\uparrow,\downarrow$ along the $\hat{z}$ direction. 
Thus, the full energy spectrum can be derived by diagonalizing separately the two sub--blocks $\mathcal{H}_{\sigma \ k}$. 
These sub--blocks are related each other, at fixed $\phi_V, \phi_{\alpha}$, by time reversal symmetry $\mathcal{T}= i \ s_y \mathcal{K}$, where $s_y$ is the Pauli matrix describing the spin operator along the $\hat{y}$ direction, and $\mathcal{K}$ is the complex conjugation operator. 
Kramers theorem then guarantees the symmetry of the energy spectrum of the full Hamiltonian $\mathcal{H}_{k}$ about $k=0$ for all $\lambda$ values.
In addition, for every periodicity $\lambda$, the energy spectrum is always twofold degenerate in the $k-$ space. For even periodicities, the energy spectrum is also symmetric about zero energy. 

A very interesting case occurs when peridicity is $\lambda=4 n$, with integer $n$, since for such periodicities the full Hamiltonian $\mathcal{H}_k$ has a chiral symmetry described by a momentum dependent operator, which gives an energy spectrum symmetric about zero energy for all $k$ values.
For instance, in the case $\lambda=4$, the chiral symmetry of the full Hamiltonian is described by the following momentum dependent operator:

\begin{eqnarray}
\mathcal{S}_{\lambda=4}= \left( 
\begin{array}{cc} 
    0              & \rho_z \\
    \rho_z e^{i k} &       0 \\
\end{array}
\right) \otimes s_x \label{chirality}
\end{eqnarray}

\noindent where \[
\rho_z= \left( 
\begin{array}{cc} 
    1              &  0\\
    0 &       -1 \\
\end{array}
\right) 
\]

\noindent and  $s_x$ is the Pauli matrix describing the spin operator along the $\hat{x}$ direction.

\begin{figure} [H]
\centering
\includegraphics[width=16cm]{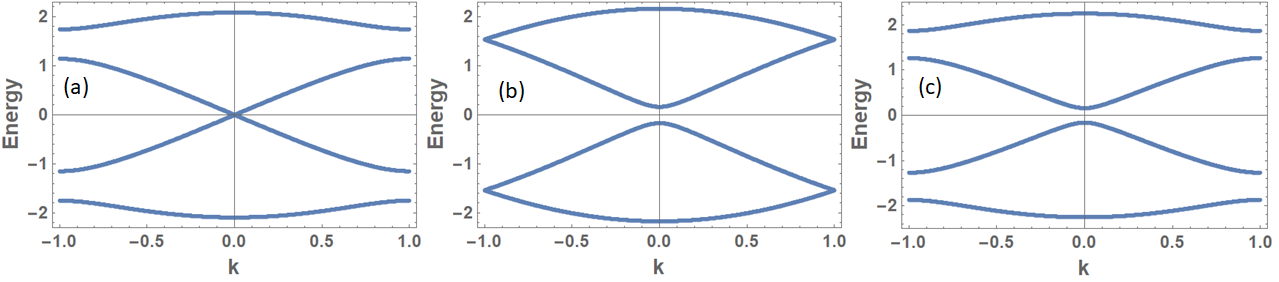}
\caption{Energy spectrum of Hamiltonian $H_k$ (Eq.(\ref{Hk})) as a function of the momentum $k$ at (a) $V_0=0.6 t, \phi_V=0$ and no RSOC, (b) $\alpha_0= 0.6 t, \phi_{\alpha}=0$ and $V_0=0$ and (c) $  V_0=\alpha_0= 0.6 t, \phi_V=0,\phi_{\alpha}=0$. All energies are measured in unit of the hopping parameter $t$.}
\end{figure}   \label{fig:spettrik}

As already demonstrated, starting from a fully metallic system, the presence of a periodic charge potential, and analogously of a periodic Rashba spin-orbit field, induces a metal-insulator transition.
However, charge potential and RSOC act in different ways, gapping out different regions of the energy spectrum, as shown in Figure 1 (a),(b), referring to the case $\lambda=4$. When only the periodic charge potential is considered (Fig.1(a)), and $\phi_V=0$, a gap opens at zone boundaries at fillings $\nu=1/4, 3/4$, while in the presence of the periodic Abelian RSOC only (with $\phi_{\alpha}=0$), the energy crossing at $k=0$ and half--filling is removed (Fig.1(b)). 
The simultaneous presence of both periodic charge potential and periodic RSOC opens gapped regions at fillings $\nu=1/4, 1/2,3/4$ (Fig.1(c)).

\begin{figure}[H]
\centering
\includegraphics[width=16cm]{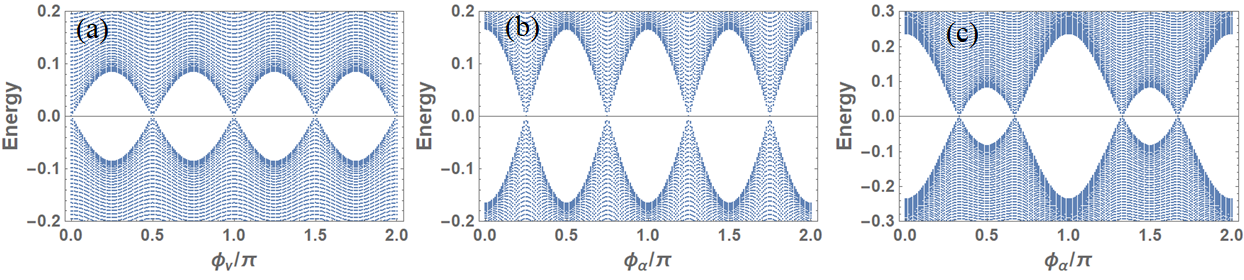}
\caption{Energy spectrum of Hamiltonian $H_k$ (a) at $V_0=0.6 t$ and no RSOC as a function of the charge potential phase $\phi_V$, (b) at $\alpha_0=0.6 t$ and no charge potential as a function of the RSOC phase $\phi_{\alpha}$, and (c) $V_0=\alpha_0=0.6 t$, $\phi_V=0.2\pi$ as a function of the RSOC phase $\phi_{\alpha}$.}
\end{figure}   \label{spettrifase}

The energy gap at half filling is very sensitive to the value of the slide parameters $\phi_V$ and $\phi_{\alpha}$, whose variations can be used to drive a gap closing and reopening in the energy spectrum at zero momentum (Fig.2).
In particular, in the presence of a periodic charge potential without RSOC, the energy crossings at half--fillings occurr when the corresponding slide phase assumes the values $\phi_V^* = m \pi/2 $ being $m$ integer as shown in Fig.2(a).
On the other side, a periodic Abelian RSOC induces an energy closing/reopening at $k=0$ when its slide value is moved across $\phi_{\alpha}^*= (m+1) \pi/4$, with integer $m$ (see Fig.2(b)). Such specific values do not depend on the amplitude of $V_0$ or $\alpha_0$.

On the other side, when the spatially modulated charge potential and RSOC both act in the system, we find that it is still possible to induce a gap closing at half filling and $k=0$, but the values of the slide phases which give rise to gap closing at half--filling strongly depend on the relative amplitude of the charge potential with respect to the RSOC. For instance, in Fig.2(c) we show the behavior of the energy spectrum of the system as a function of $\phi_{\alpha}$ at $\lambda=4$ for $V_0=\alpha_0=0.6 t$ and $\phi_V=0.2 \pi$. As it is possible to notice, the energy at half--filling closes at values of $\phi_{\alpha}$ which do not correspond neither to the slide phase values for the limit case of RSOC only or to those of the case with periodic charge potential only. 

\begin{figure} 
\centering
\includegraphics[width=16cm]{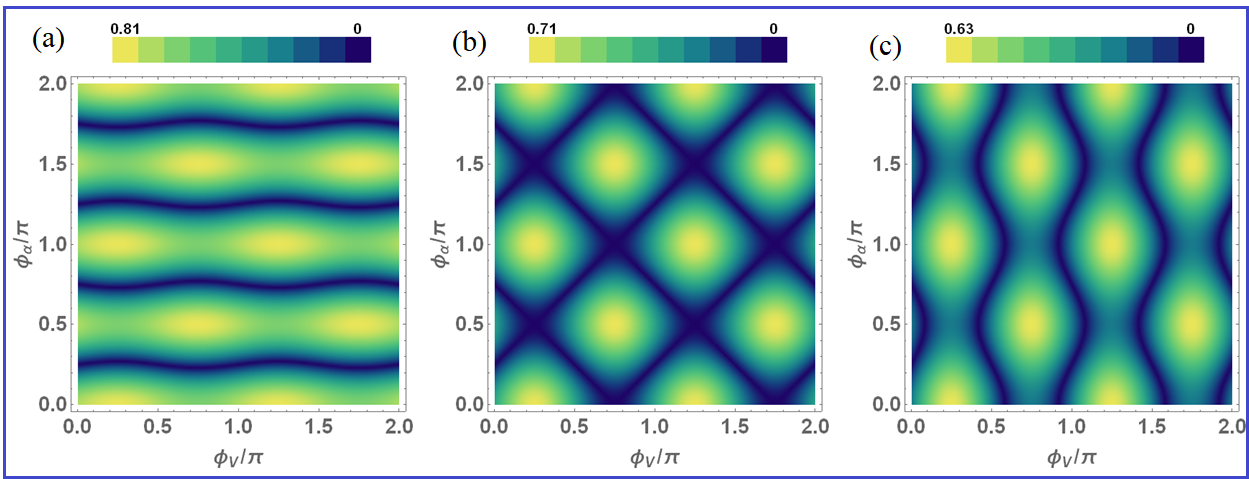}
\caption{Density plot of the energy gap between the 4th and the 5th energy bands (counted starting from the lowest one in energy) for periodicity $\lambda =4$ at $k=0$ in the synthetic space $\left( \phi_V, \phi_{\alpha}\right) $ at $\alpha_0= t$ and $V_0=\alpha_0/2$ in (a),$V_0=\sqrt{2}\alpha_0$ in (b) and $V_0=2 \alpha_0$ in (c). }\label{Nodallines}
\end{figure}   

Thus, if we analyze the energy gap at half-filling in the synthetic plane $\left( \phi_{V},  \phi_{\alpha}\right)$, we find that energy assumes a nematic nodal structure in this plane, such that the topology of the energy nodal lines strongly depends on the ratio $V_0/\alpha_0$, as it is possible to see in Fig.\ref{Nodallines}, where we have shown a density plot of the energy gap at half filling for the case $\lambda=4$ as a function of the two slide phases $\ \phi_{V},  \phi_{\alpha}$ for three different values of the ratio $V_0/\alpha_0$: $V_0/\alpha_0 =0.5$ in (a), $V_0/\alpha_0=\sqrt{2}$ in (b), and $V_0/\alpha_0=2$ in (c). 
For values of the ratio $V_0/\alpha_0$ lower that $\sqrt{2}$, there are almost horizontal stripes in the plane $\left(\phi_{V},  \phi_{\alpha} \right)$. They have insulating character and are separated each other by nodal lines which are in turn mainly constant as a function of $\phi_{\alpha}$.

Moving towards $V_0/\alpha_0=\sqrt{2}$ the insulating stripes, as well as the nodal lines, gradually deform, untill they merge forming a checkerboard-like structure of confined squared insulating plaquettes. By further increasing the ratio $V_0/\alpha_0$, the stripe structure starts to reappear, but rotated by $90\degree$.

\begin{figure}[H]
\centering
\includegraphics[width=16cm]{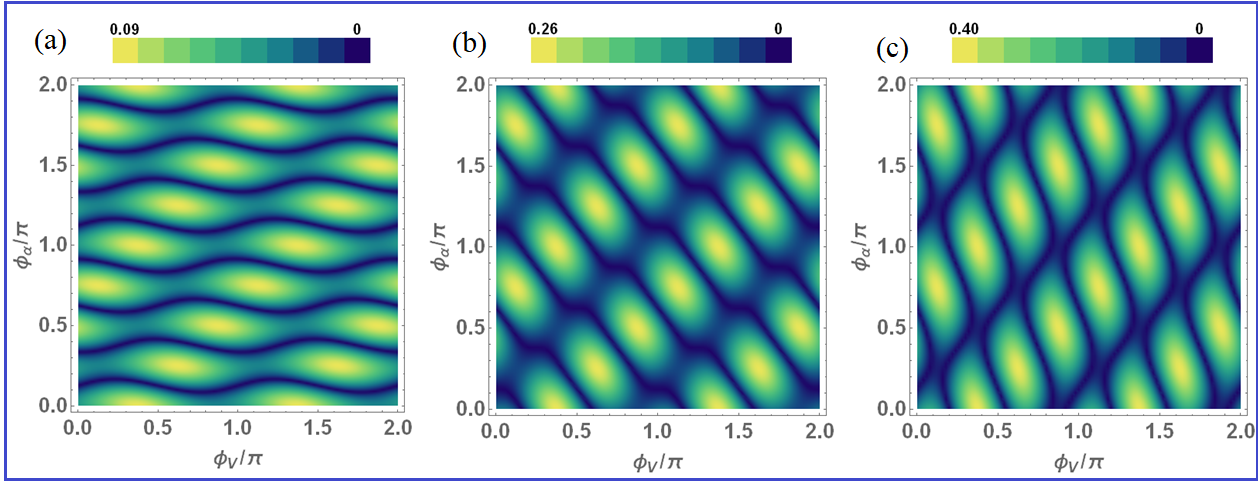}
\caption{Density plot of the energy gap between the 8th and the 9th bands (counted starting from the lowest one in energy) for periodicty $\lambda =8$ at $k=0$ in the syntetic space $\left( \phi_V, \phi_{\alpha}\right) $ at $\alpha_0= t$ and $V_0=\alpha_0/2$ in (a),$V_0=\sqrt{2}\alpha_0$ in (b) and $V_0=2 \alpha_0$ in (c). }\label{Nodallines8}
\end{figure}   

Such "nematic" character only depends on the relative value $V_0/\alpha_0$, and can be observed at all periodicity values $\lambda=4 n$, for integer values of $n$. The number of nodal lines, and correspondingly the number of separated insulating stripes in the synthetic plane $\left( \phi_{V},  \phi_{\alpha}\right)$ scales with the periodicity $\lambda$ and it is actually given by the value of the integer $n$, as it is possible to see in Fig.\ref{Nodallines8}, which shows the density plot of the energy gap at half filling and $k=0$ for periodicity $\lambda=8$.

We have verified that for other periodicities $\lambda= 4 n$ such nodal structure still holds.

\section{Symmetry protection of the nodal lines}
 
The symmetry analysis of the full Hamiltonian $H_k$ shows that each spin sub-matrix $\mathcal{H}_{\sigma k}$ breaks time-reversal symmetry while preserving particle–hole symmetry. 
In the case of $\lambda=4$, the particle-hole operator is

\begin{eqnarray} 
\mathcal{C}_k= \left( 
\begin{array}{cc} 
  0  &  e^{- i k/2} \rho_z \\
   e^{i k/2} \rho_z &   0  \\
\end{array}
\right) \mathcal{K},
\end{eqnarray}

\noindent and it is such that $\mathcal{C}_{k}^2=1$ for all $k$ values and
\[\mathcal{C}_{k}^{-1} \mathcal{H}_{\sigma k} (\phi_V, \phi_{\alpha}) \mathcal{C} + \mathcal{H}^{*}_{-\sigma k} (\phi_V, \phi_{\alpha})=0
\]

\noindent being $\mathcal{A}^{*}$ the complex conjugate of a generic operator $A$.

Thus, in the plane $\left(\phi_V,\phi_{\alpha}\right)$ at $k=0$, the system behaves like a two-dimensional nodal superconductor. 
According to the classification of gapless topological phases \cite{classificationnodal}, this is a class D system with codimension 1, which thus allows a topological phase characterized by a $Z_2$ topological index.

\begin{figure}[H]
\centering
\includegraphics[width=16cm]{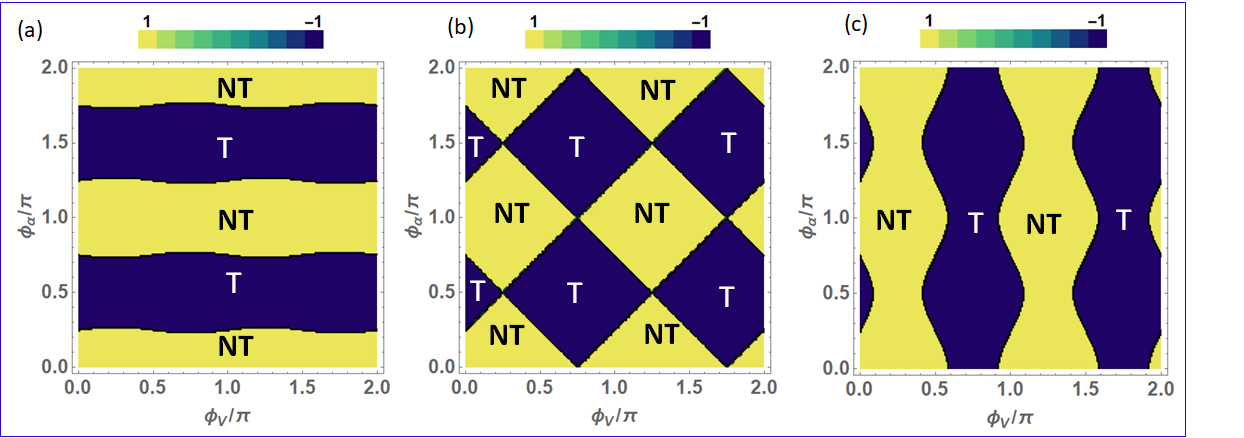}
\caption{Density plot of the topological invariant and corresponding classification of the different regions as topological (T) and non topological (NT) at $\alpha_0= t$ and $V_0=\alpha_0/2$ in (a),$V_0=\sqrt{2}\alpha_0$ in (b) and $V_0=2 \alpha_0$ in (c).} \label{pf}
\end{figure}   

Therefore, we can characterize the insulating regions found in the plane  $\left(\phi_V,\phi_{\alpha}\right)$ at $k=0$ in terms of the fermion parity $\mathcal{P}$ of the ground state \cite{Loring}, i.e., as the sign of the Pfaffian of the Hamiltonian in Majorana representation. The fermion parity labels the topological inequivalent ground states: the trivial state has $\mathcal{P}=1$ (even parity) while nontrivial state  has $\mathcal{P}=-1$ (odd parity).

The fermion parity of the topological quantum system described by the Hamiltonian  $H_{\sigma 0}$ can be evaluated analytically, as $\mathcal{P}=Sign\left[\sqrt{(Det(\mathcal{C}_0   H_{\sigma 0})}\right]$.  The points where the fermion parity changes from $+1$ to $-1$ and viceversa correspond to the gapless lines where zero-energy states occurr. By performing the analysis of the fermion parity for the case $\lambda=4$, we established the topological character of the insulating regions of the energy spectrum at $k=0$ and half--filling, as shown in Fig.\ref{pf}, for three different values of the ratio $V_0/\alpha_0$.  
 
We would like to point out that the zero energy states driven by the slide phase of a single periodic field in the two limit cases (periodic charge field only or periodic RSOC only), have a topological origin totally different than that of the zero energy states in the case of periodic charge potential and RSOC together. Indeed, although the occurrence of the energy crossings for the periodic charge potential only and for the periodic Abelian RSOC only happens at different slide values, in both cases their origin can be understood in terms of the symmetries characterizing the corresponding Hamiltonian matrix sub-blocks $\mathcal{H}_{ \sigma \ k}$.

Indeed in both cases, at the phases values which give the energy gap closing at $k=0$ and half--filling, we found that $\mathcal{H}_{\sigma \ k=0}$ is characterized by a chiral $\mathcal{S}$ and a mirror $\mathcal{M}$ symmetry, which are represented by operators commuting each other. In such a case, it is possible to determine the possible presence of zero energy states in the energy spectrum through the definition of a topological invariant, given by the sum of the trace of the chirality sub-blocks labelled by the different mirror eigenvalues \cite{Sato2014}. In the both cases of periodic charge field only and periodic RSOC only we find that this topological number is finite only at the slide phase values where energy crossings at half--fillings happen, thus demonstrating that they are topologically protected by chirality and mirror symmetries.

\section{Conclusions}  
We have analyzed the combined effect of the application of modulated charge potential and modulated Rashba spin-orbit coupling on a metallic atomic chain. We have demonstrated a feasible way to drive the system into a topological nodal line semimetal by sliding the phases of charge potential and RSOC, for periodicity values $\lambda= 4 n$ with integer $n$, in the half-filling regime. The topological insulating regime we find is characterized by topological insulating stripes, separated by non-topological ones through nodal lines within the syntetic space $(\phi_V, \phi_{\alpha})$. The resulting pattern has a "nematic" structure, which strongly depends on the relative amplitude of the charge potential and RSOC field. Indeed, by tuning this parameter it is possible to modify the structure of the nodal lines as well as the topological phase transitions in the space $(\phi_V, \phi_{\alpha})$, and thus change on demand the shape and the extension of the area in the synthetic space where the topological phase exists. 
The emerging topological nodal line semimetallic phase is protected by particle-hole symmetry. Our results show an interesting and simple way to dynamically  switch on and off  the topological insulating phases by exploiting spacially modulated charge potential and RSOC. 
  
Physical platforms where such possibilities can be realized and tested  are represented by Fermi gases loaded in periodic optical lattice, as well as in semiconductor nanowires with perpendicular modulated voltage gates, or with complex geometrical shapes, obtainable via nanostructuring methods like for instance electron beam lithography or adhesion of ribbons on prestrained substrates.
Within this context, low--dimensional systems with non trivial geometry have been already demonstrated to offer a potential playgroung for triggering new  functionalities through the exploitation of curvature effects, like for instance in tuning the electron spin interference \cite{PRBrings} and the superconducting state \cite{PRBSC} in closed loop configurations, as well as the supercurrent in weak links between Rashba coupled superconducting nanowires with geometric misalignment \cite{ISEC}. Our findings therefore add a new important piece to the rich puzzle of unique curvature-induced quantum effects in low-dimensional semiconducting systems.  
\vspace{6pt} 



\authorcontributions{``Conceptualization, Paola Gentile, Carmine Ortix, Canio Noce and Mario Cuoco; Data curation, Paola Gentile and Vittorio Benvenuto; Formal analysis, Paola Gentile and Mario Cuoco; Investigation, Paola Gentile, Vittorio Benvenuto, Carmine Ortix, Canio Noce and Mario Cuoco; Methodology, Paola Gentile and Mario Cuoco; Writing -- original draft, Paola Gentile and Mario Cuoco; Writing -- review and editing, Vittorio Benvenuto, Carmine Ortix, Canio Noce''}

%
 \acknowledgments{C.O. acknowledges support from a VIDI grant (Project 680-47-543) financed by the Netherlands Organization for Scientific Research (NWO)}

\conflictsofinterest{``The authors declare no conflict of interest.''}

\reftitle{References}



\begin{thebibliography}{999}

\bibitem{art1} Bernevig, B. A.; Hughes, T. L.; Zhang, S.-C. Quantum Spin Hall Effect and Topological Phase Transition in HgTe Quantum Wells. {\em Science} {\bf 2006}, {\em 314}, 1757--1761. 

\bibitem{art2}  K\"{o}nig, M.; Wiedmann, S.;  Br\"{u}une, C.; Roth, A.;  Buhmann, H.;  Molenkamp, L. W. ;  Qi, X.-L.; Zhang, S.-C. Quantum Spin Hall Insulator State in HgTe Quantum Wells. {\em Science} {\bf 2007}, {\em 318}, 766--770.

\bibitem{art3}  K\"{o}nig, M.; Buhmann, H.;  Molenkamp, L. W.; Hughes, T.;  Liu, C.-X. ; Qi, X.-L.; Zhang, S.-C. The Quantum Spin Hall Effect: Theory and Experiment. {\em J. Phys. Soc. Jpn.}{\bf 2008}, {\em 77}, 031007--031021. 

\bibitem{art4}  Roth, A.; Br\"{u}ne, C. ; Buhmann,  H. ; Molenkamp, L. W.;   Maciejko, J.; Qi, X.-L.; Zhang,  S.-C. Nonlocal Transport in the Quantum Spin Hall State. {\em Science}{\bf 2009}, {\em 325}, 294--297. 

\bibitem{art5}  Xia, Y.; Qian, D. ; Hsieh,  D.;  Wray, L.; Pal,  A. ; Lin, H. ;  Bansil, A.; Grauer,  D.; Hor, Y. S.; Cava, R. J.; Hasan,  M. Z. Observation of a large-gap topological-insulator class with a single Dirac cone on the surface. {\em Nature Physics}{\bf 2009}, {\em 5}, 398--402. 

\bibitem{art6} Zhang, H.; Liu, C.-X.; Qi,X. -L.; Dai, X.; Fang, Z.; Zhang, S.-C.  Topological insulators in Bi$_2$Se$_3$, Bi$_2$Te$_3$ and Sb$_2$Te$_3$ with a single Dirac cone on the surface. {\em Nature Physics} {\bf 2009}, {\em 5}, 438--442. 

\bibitem{art7}  Jiang,  H.;  Cheng, S. G.; Sun, Q. F. ; Xie, X. C.  Topological Insulator: A New Quantized Spin Hall Resistance Robust to Dephasing. {\em Phys. Rev. Lett.}{\bf 2009}, {\em 103}, 036803-1--036803-4.

\bibitem{art8} Hasan, M. Z.; Kane, C. L. {\it Colloquium}: Topological insulators. {\em Review of Modern Physics} {\bf 2010}, {\em 82}, 3045--3067.

\bibitem{art9} Qi, X.-L.;  Zhang, S.-C.  Topological insulators and superconductors {\em Review of Modern Physics.} {\bf 2011}, {\em 83}, 1057--1110.

\bibitem{art10} Qiao, Z. H. ; Tse, W. K.; Jiang, H.;  Yao,  Y. G.;  Niu, Q. 
Two-Dimensional Topological Insulator State and Topological Phase Transition in Bilayer Graphene. {\em Physical Review Letters} {\bf 2011}, {\em 107}, 256801-1 -- 256801-4.

\bibitem{art11} Chang,  C. Z.;  Zhang, J. S.; Feng,  X.;  Shen, J.; Zhang,   Z. C.;  Guo, M. H.;   Li, K.; Ou,   Y. B.; Wei,  P.;  Wang, L. L.; Ji, Z. Q.;  Feng, Y. ; Ji, S. H.;  Chen, X. ; Jia, J. F. ; Dai,  X.; Fang, Z.; Zhang, S.-C.; He, K. ; Wang, Y. Y. ; Lu, L.; Ma, X. C.; Xue, Q. K. Experimental Observation of the Quantum Anomalous Hall Effect in a Magnetic Topological Insulator. {\em Science} {\bf 2013}, {\em 340}, 167--170.

\bibitem{art12} Wang, Z. F.; Liu,  Z. ; Liu, F. Quantum anomalous Hall effect in 2D organic topological insulators. {\em Phys. Rev. Lett.} {\bf 2013}, {\em 110}, 196801-1 --  196801-4.

\bibitem{art13}  Thouless, D. J.; Kohmoto, M. ; Nightingale, M. P.; den Nijs, M. Quantized Hall Conductance in a Two-Dimensional Periodic Potential. {\em Phys. Rev. Lett.} {\bf 1982}, {\em 49}, 405--408 . 

\bibitem{art14} Kane,  C. L. ; Mele, E. J.  Z$_2$ Topological Order and the Quantum Spin Hall Effect. {\em Phys. Rev. Lett} {\bf 2005}, {\em 95}, 146802-1 -- 146802--2 (2005). 

\bibitem{art15}  Sheng, D. N.; Weng, Z. Y.; Sheng, L.; Haldane, F. D. M. Quantum Spin-Hall Effect and Topologically Invariant Chern Numbers. {\em Phys. Rev. Lett.} {\bf 2006}, {\em 97}, 036808-1 --036808-4.

\bibitem{art17} Li, J.; Chu, R. L.; Jain, J. K.; Shen, S. Q. Topological Anderson Insulator. {\em Phys. Rev. Lett.} {\bf 2009}, {\em 102}, 136806-1 -- 136806-4.  

\bibitem{art18} Jiang,  Z. F.; Chu, R. L.; Shen, S. Q. Electric-field modulation of the number of helical edge states in thin-film semiconductors. {\em Phys. Rev. B} {\bf 2010}, {\em 81}, 115322-1 -- 115322-4.

\bibitem{art19} Kim, M.; Kim, C. H.; Kim,  H. S.; Ihm, J. Topological quantum phase transitions driven by external electric fields in Sb$_2$Te$_3$ thin films. {\em Proc. Natl. Acad. Sci. USA} {\bf 2012}, {\em 109}, 671-674.

\bibitem{art20} Bahramy,  M. ;  Yang, B. J. ; Arita, R.; Nagaosa, N. Emergent quantum confinement at topological insulator surfaces. {\em Nat. Commun.}{\bf 2012} {\em 3}, 1159,1--7.

\bibitem{art21} Esaki L. ; Tsu, R. Superlattice and negative differential conductivity in semiconductors. {\em IBM J. Res. Develop.} {\bf 1970}, {\em 14}, 61-65; Tsu, R. {\em Superlattice to Nanoelectronics}; Elsevier: Oxford, 2005.

\bibitem{art22} Schneider, H. .;  Grahn, H. T. ; Klitzing,  K. V.;  Ploog, K.  Sequential resonant tunneling of holes in GaAs-AlAs superlattices. {\em Phys. Rev. B} {\bf 1989}, {\em 40}, 10040-10043(R). 
%
\bibitem{art23} Wacker, A.; Moscoso, M.; Kindelan,  M. ; Bonilla,  L. L.  Current-voltage characteristic and stability in resonant-tunneling n-dopedsemiconductor superlattices. {\em  Phys. Rev. B} {\bf 2003}, {\em 55}, 2466-2475. 
%
 \bibitem{art24} Cheng, Y.-C.; Yang, S.-T.; Yang, J.-N.; Lan,  W.-H.; Chang,  L.-B.;  Hsieh, L.-Z.  Fabrication of far-infrared photodetector based on InAs/GaAs quantum dot superlattices.  {\em Opt. Eng.}{\bf 2003} 42(1), 119-123.
 
\bibitem{PhysicaB} Zheng, Y.; Yang, S.-J. Topological bands in one-dimensional periodic potentials. {\em Physica B} {\bf  2014}, {\em 454}, 93-97.

\bibitem{PRB2014} Fu, B.; Zheng, H.; Li, Q.; Shi, Q., Yang, J.  Topological phase transition driven by a spatially periodic potential. {\em Phys. Rev. B} {\bf 2014}, {\em 90}, 214502-1 -- 214502-6.

\bibitem{nostroPRL2015}  Gentile, P.; Cuoco, M.; Ortix, C. Edge States and Topological Insulating Phases Generated by Curving a Nanowire with Rashba Spin-Orbit Coupling. {\em Phys. Rev. Lett.}{\bf 2015}, {\em 115}, 256801.

\bibitem{PRBserpentine}  Pandey, S.; Scopigno, N.; Gentile, P.; Cuoco, M.; Ortix, C.  Topological quantum pump in serpentine-shaped semiconducting narrow channels. {\em  Phys. Rev. B} {\bf 2018}, {\em 97}, 241103-1 --241103-5 (R).

\bibitem{OrtixLau2016}  Lau, A., van den Brink,  J.; Ortix, C. Topological mirror insulators in one dimension. {\em Phys. Rev. B} {\bf 2016}, {\em 94} 165164-1 -- 165164-9  

\bibitem{AAH1} Harper, P. G. Single Band Motion of Conduction Electrons in a Uniform Magnetic Field. {\em  Proc. Phys. Soc. London Sect. A} {\bf 1955}, {\em 68}, 874--878.

\bibitem{AAH2} Aubry, S. and Andr\'{e}, G.  Analyticity breaking and Anderson localization in incommensurate lattices. {\em Ann. Isr. Phys. Soc.}{\bf 1980}, {\em 3}, 133--151.

\bibitem{AAH3} Ganeshan, S.;  Sun, K;  and Das Sarma, S. Topological zero-energy modes in gapless commensurate Aubry-André-Harper models. {\em Phys. Rev. Lett.} {\bf 2013} {\em 110}, 180403-1 --180403-5.


\bibitem{classificationnodal} 
Matsuura, S.; Chang, P.-Y. ; Schnyder A. P.; Ryu, S. Protected boundary states in gapless topological phases.  {\em New Journal of Physics} {\bf 2013}, {\em 15}, 065001-1 --065001-25 

\bibitem{Loring}  Loring, T. A. {\it K}-theory and pseudospectra for topological insulators. {\em Ann. Phys.} {\bf 2015}, {\em 356}, 383–416.

\bibitem{Sato2014} Koshino, M. , Morimoto, T.; Sato, M.  Topological zero modes and Dirac points protected by spatial symmetry and chiral symmetry. {\em Phys. Rev. B} {\bf 2014}, {\em 90}, 115207-1 -- 115207-11.

\bibitem{PRBrings} Ying, Z.-J. ; Gentile, P.; Ortix, C.; Cuoco, M. Designing electron spin textures and spin interferometers by shape deformations. {\em Phys. Rev. B}{\bf 2016} {\em 94}, 081406-1 --081406-5 (R). 

\bibitem{PRBSC}  Ying, Z. Y., Cuoco, M., Ortix, C. ; Gentile, P. Tuning pairing amplitude and spin-triplet texture by curving superconducting nanostructures. {\em Phys. Rev. B}{\bf 2017}, {\em 96}, 100506-1 -- 100506-6 (R).

\bibitem{ISEC}  Ying, Z.-J.; Cuoco, M.; Gentile, P.; Ortix, C.  Josephson Current in Rashba-Based Superconducting Nanowires with Geometric Misalignment: Rashba-Based Superconducting Nanowires with Geometric Misalignment. In {\em  2017 16th International Superconductive Electronics Conference (ISEC) }; IEEE, New York, 2017, pp. 1-–3.  
 
\end{thebibliography}




\end{document}